\documentclass[aps,pra, twocolumn,a4paper,showpacs]{revtex4}
\usepackage{amssymb,amsmath,amsfonts,graphicx}
\usepackage{bm}

\def\BE {\begin{equation}}
\def\EE {\end{equation}}
\def\BEA {\begin{eqnarray}}
\def\EEA {\end{eqnarray}}
\def\BES {\begin{subequations}}
\def\EES {\end{subequations}}
\def\BA {\begin{array}}
\def\EA {\end{array}}

\def\ep {\epsilon}

\def\ef{^{{\rm e}}}

{

\begin{document}

\title{Optimized time-dependent perturbation theory for pulse-driven quantum 
dynamics in atomic or molecular systems}
\date{\today}
\author{D.~Daems}
\email{ddaems@ulb.ac.be}
\affiliation{Center for Nonlinear Phenomena and Complex Systems, Universit\'e Libre de
Bruxelles, CP 231, 1050 Brussels, Belgium}
\author{S.~Gu\'erin}
\affiliation{Laboratoire de Physique de l'Universit\'e de Bourgogne, UMR CNRS 5027, BP
47870, 21078 Dijon, France}
\author{H. R.~Jauslin}
\affiliation{Laboratoire de Physique de l'Universit\'e de Bourgogne, UMR CNRS 5027, BP
47870, 21078 Dijon, France}
\author{A.~Keller}
\affiliation{Laboratoire de Photophysique Mol\'eculaire du CNRS, Universit\'e Paris-Sud,
B\^at. 210 - Campus d'Orsay, 91405 Orsay Cedex, France}
\author{O.~Atabek}
\affiliation{Laboratoire de Photophysique Mol\'eculaire du CNRS, Universit\'e Paris-Sud,
B\^at. 210 - Campus d'Orsay, 91405 Orsay Cedex, France}

\begin{abstract}
We present a time-dependent perturbative approach adapted to the treatment
of intense pulsed interactions. 
We show there is a freedom in choosing
secular terms and use it to optimize the accuracy of the
approximation. We apply this formulation to a unitary superconvergent
technique and improve the accuracy by several orders of magnitude with
respect to the Magnus expansion.
\end{abstract}

\pacs{31.15.Md, 03.65.-w, 42.50.Hz}
\maketitle



Perturbation theory when combined with a specific
treatment for resonances is quite well understood in classical and quantum
mechanics for time-independent systems. This includes also time-periodic
driven systems for which the periodicity can be treated by Floquet theory in
a way that yields a time-independent formulation \cite{Barata,Adv}. One
knows that resonances yield divergent terms, that appear as small
denominators, which have to be specifically removed. The counterpart of the
concept of resonance for time-dependent systems is generally associated to
secular terms whose size grows with time (see Ref.~\cite{Langhoff} and references
therein).

With the advent of short ($\simeq 10$~ fs) and intense ($10^{13}$--$10^{15}$%
~W/cm$^{2}$) laser pulses, atomic or molecular systems can be strongly
perturbed in a timescale shorter than characteristic times corresponding to
the free evolution of the system and adiabatic theories are not applicable
(see, e.g., Ref.~\cite{align}). The goal of this paper is to formulate a
time-dependent perturbation theory well adapted for perturbations localized in time.

The conceptual framework of perturbation theory can be described as follows:
The Hamiltonian of the considered system can be decomposed as the sum of two
terms $H_{1}=H_{0}+\ep V_{1}$. The first term $H_{0}$ is assumed to have a structure simple enough to lead to
explicitly known solutions for its associated propagator $U_{H_0}(t, t_0)$. 
The term $\ep V_{1}$ is supposed to be
small with respect to $H_{0}$, in a sense specified below.
Time-independent perturbation theories can be equivalently formulated at the
level of eigenvectors or operators \cite{Primas}. A large class of these approaches amounts to construct a unitary transformation $T$ such that%
\begin{equation}
T^{\dagger }H_{1}T=H\ef+\ep^{\prime}V^{\prime }\text{,}
\end{equation}%
where $H\ef$ is still of simple structure [i.e., its propagator $%
U_{H\ef}(t,t_{0})$ can be explicitly computed] and $\ep^{\prime}V^{\prime
}$ is a perturbation whose size is smaller than the original one. 
To compute the transformation $T$ explicitly, one represents it in general either (i) in terms of some power
series 
\begin{equation}
T=e^{-iW},\quad W=\sum_{k}\ep^{k}W_{k} , \label{recur}
\end{equation}%
or (ii) by an iterative construction as a composition of
transformations 
\begin{equation}
T=\prod_{k}e^{-i\ep_{k}W_{k}}.  \label{it}
\end{equation}%
These procedures generally differ.
The former one is referred to as the
time-independent Poincar\'{e}-Von Zeipel technique, which has been shown to
be equivalent to the usual Rayleigh-Schr\"{o}dinger perturbation theory \cite%
{SchererI}.
The latter procedure includes  the Van Vleck technique (for which $\ep_k=\ep^k$) and the superconvergent Kolmogorov-Arnold-Moser (KAM) expansion (where $\ep_k=\ep^{2^{k-1}}$ and $W_k$ is $\ep_k$-dependent) %
\cite{Scherer95}. The perturbative procedure converges if the remaining
perturbation $\ep^{\prime}V^{\prime }$ can be made to go to zero, as the number of terms
in the power series (\ref{recur}) or as the number of
compositions in Eq.~(\ref{it}) goes to infinity.

In this description one has to state precisely what class of Hamiltonians $%
H\ef$ can be considered simple. For the first order or the first
iteration, one considers $H\ef=H_{0}+\ep D_{1}$ with the
condition that $D_{1}$ should be \textit{compatible} with $H_{0}$ in the
sense that if the propagator of $H_{0}$ is known, that of $H_{0}+\ep %
D_{1}$ can also be obtained explicitly. In the case of time-independent
Hamiltonians the condition of compatibility is%
\begin{equation}
\left[ H_{0},D_{1}\right] =0.  \label{DDD1}
\end{equation}%
For the case of time-dependent Hamiltonians, we
show that the condition of compatibility can be generalized to%
\begin{equation}
\left[ H_{0}(t),D_{1}(t)\right] =i\frac{\partial D_{1}}{\partial t}.
\label{DD1}
\end{equation}%
The construction of transformations of the type of Eq. (\ref{recur}) or (%
\ref{it}) involves finding the generator $-i\ep W_{1}(t)$ of the transformation $T_1(t)$, that
is  the solution to
\begin{equation}
i\left[ W_{1}(t),H_{0}(t)\right] +V_{1}(t)-D_{1}(t)=\frac{\partial W_{1}}{%
\partial t}.  \label{DW1}
\end{equation}%
This equation, together with the constraint (\ref{DD1}), are usually called \textit{%
cohomology equations} in the time-independent case \cite{PhysicaA} and are
here generalized to the time-dependent case. These cohomology equations have the same form for higher orders or successive iterations.

Here we formulate the time-dependent perturbation theory by
transforming directly the evolution
operator instead of considering the perturbed Hamiltonian as is usually done in time-independent theory. 
We obtain perturbative corrections to the full propagator in the form of a product of propagators which exhibit free parameters appearing through the general solutions of related differential equations.
We recover in particular the Magnus expansion \cite{Pechukas} as a special case of the
time-dependent Poincar\'{e}-Von Zeipel theory. 
This extension also gives the
precise correspondence between time-independent resonances and
time-dependent secular terms. 
In the context of  pulsed 
perturbations with a finite duration, the secular terms need not be eliminated. We
show the remarkable result that they can be used to improve the convergence
of the method at a given order.
This optimization is achieved 
without any {\em a priori} knowledge of the
solution by locating the minimum of a given eigenvalue as a function of the relevant free parameters that are identified. 
The efficiency of the method is illustrated on a two-level system
driven by a short intense pulse.

{\it Perturbation theory, resonances and secular terms}.
We consider the
Hamiltonian $H_{1}(t)=H_{0}(t)+\ep V_{1}(t)$, where $H_{0}(t)$ is associated
to a known propagator $U_{H_{0}}(t,t_{0})$.
The formulation is presented here for the  KAM method, consisting in
iterations of transformations which are exactly of the same form at each step. 
The first iteration involves a unitary operator $T_{1}(t)$ which transforms the propagator $%
U_{H_{1}}(t,t_{0})$ according to
\begin{equation}
T_{1}^{\dagger }(t)U_{H_{1}}(t,t_{0})T_{1}(t_{0})=U_{H_{2}}(t,t_{0}),
\end{equation}%
into a propagator $U_{H_{2}}(t,t_{0})$ associated with the sum $H_2(t)$ of an effective
Hamiltonian $H_{1}^{\mathrm{e}}(t)\equiv H_{0}(t)+\ep D_{1}(t)$ which
contains contributions up to order $\ep$ and a remainder $\ep^{2}V_{2}(t)$. 
This new propagator, generated by
a sum of two Hamiltonians, can be written as the product%
\begin{equation}
U_{H_{2}}(t,t_{0})=U_{H_{1}^{\text{e}}}(t,t_{0})R_{2}(t,t_{0}),  \label{R2}
\end{equation}%
where $R_{2}(t,t_{0})$   satisfies the Schr\"odinger equation whose Hamiltonian is $%
\ep^{2}U_{H_{1}^{\text{e}}}(t_{0},t)V_{2}(t)U_{H_{1}^{\text{e}}}(t,t_{0})$.
Similarly we can factorize 
$U_{H_{1}^{\text{e}}}(t,t_{0})=U_{H_{0}}(t,t_{0})S_{1}(t,t_{0})$, where $%
S_{1}(t,t_{0})$ is a unitary operator related to $\ep D_{1}(t)$: 
\begin{equation}
i\frac{\partial }{\partial t}S_{1}(t,t_{0})=\ep %
U_{H_{0}}(t_{0},t)D_{1}(t)U_{H_{0}}(t,t_{0})\,S_{1}(t,t_{0}).  \label{S1}
\end{equation}%
The full propagator reads 
\BEA
\label{UH1UH2}
U_{H_{1}}(t,t_{0})=T_{1}(t)U_{H_{0}}(t,t_{0})S_{1}(t,t_{0})R_{2}(t,t_{0})T_{1}^{\dagger }(t_{0}),\quad
\EEA%
which yields the first order KAM approximation  for $R_{2}(t,t_{0})$ replaced by
the identity. 
In this construction the only restriction on the self-adjoint operator $%
\ep D_{1}(t)$ is that it be of order $\ep$. Hence we have the freedom to
choose the Hamiltonian in Eq. (\ref{S1}) as $t$-independent, giving%
\begin{equation}
D_{1}(t)=U_{H_{0}}(t,t_{0})D_{1}(t_{0})U_{H_{0}}(t_{0},t) , \quad \label{DD1(t)}
\end{equation}%
with $D_{1}(t_{0})$ arbitrary, which is the general solution of Eq. (%
\ref{DD1}). This allows one to obtain the solution of Eq. (\ref%
{S1}) as%
\begin{equation}
S_{1}(t,t_{0})=\exp \left[ -i(t-t_{0})\ep D_{1}(t_{0})\right] .
\end{equation}%
Differentiating Eq.~(\ref{UH1UH2}) and substituting $T_{1}(t)=\exp \left( -i\ep W_{1}(t)\right)$ leads to Eq.~(\ref{DW1}).   The
rest involves a series of $k$ nested commutators that reads  
$
\ep^2 V_{2}(t)=\sum_{k=1}^{\infty }\frac{1}{(k+1)!}i^k\ep^{k+1}\, 
\left[W_1(t), \cdots\left[W_1(t),kV_{1}(t)+D_{1}(t)\right]\cdots\right]$. It has exactly the same structure at each iteration which is
useful for applications, particularly when high-order computations are
needed.

Iterating the time-dependent KAM algorithm reduces the size of the remaining perturbation in a
superconvergent way from order $\ep^{2^{n-1}}$ to $\ep^{2^{n}}$ at step $n$. 
The time-dependent Van Vleck technique would allow one to reduce the
size of the remainder from order $%
\ep^{n}$ to $\ep^{n+1}$. 
These methods, in the formulation presented here, are unitary upon truncation. 
The superconvergent character of the KAM algorithm has been shown numerically by
applying the method to a two-level system perturbed by a short time-dependent interaction \cite{art1}.

The \textit{time-independent problem}, i.e., the problem of finding a
transformation $T_{1}$ that enables one to simplify the time-independent
Hamiltonian $H_{1}$ according to $T_{1}^{\dagger }HT_{1}=H_{0}+\ep D_{1}+\ep%
^{2}V_{2}$, is recovered when one conveniently chooses $%
T_{1}$ as time-independent. 
In this case all the operators, and in
particular $D_{1}$ and $W_{1}$, are
time-independent and the standard cohomology equations are recovered: $\left[
H_{0},D_{1}\right] =0$ and $V_{1}-D_{1}+i\left[ W_{1},H_{0}\right] =0$. Their
solutions can be determined using the
following key property \cite{PhysicaA}: $W_{1}$ exists if and only if $\Pi
_{H_{0}}(D_{1}-V_{1})=0$, where $\Pi _{H_{0}}$ is the projector in the
kernel of the application $A\mapsto \lbrack A,H_{0}]$ (for an operator $A$
acting on the same Hilbert space as $H_{0}$). The projector $\Pi _{H_{0}}$
applied on an operator $A$ captures thus all the part $B$ of $A$ which
commutes with $H_{0}$: $[B,H_{0}]=0$. The unique solution $D_{1}$ allowing $%
W_{1}$ to exist and satisfying Eq. (\ref{DDD1}) is thus%
\begin{equation}
D_{1}=\Pi _{H_{0}}V_{1}\equiv \lim_{T\rightarrow \infty }\frac{1}{T}%
\int_{0}^{T}e^{-itH_{0}}V_{1}e^{itH_{0}}.  \label{D1}
\end{equation}%
The \textit{resonances} are associated with terms of $V_{1}$ which commute
with $H_{0}$. Application of Eq. (\ref{D1}) can be interpreted as an \textit{%
averaging }of $V_{1}$ with respect to $H_{0}$ which allows one to extract resonances.

For the \textit{time-dependent problem}, the general solution of Eq. (\ref%
{DW1}) reads (up to a term $U_{H_0}(t,t_0)B_1 U_{H_0}(t_0,t)$ with $B_1$ any self-adjoint operator that we set here to 0)%
\begin{eqnarray}
W_{1}(t)=\int_{t_{1}^{\prime }}^{t}dsU_{H_{0}}(t,s)\left(
V_{1}(s)-D_{1}(s)\right) U_{H_{0}}(s,t)  , \quad
  \label{W1(t)}
\end{eqnarray}%
with  $t_{1}^{\prime }$ any real number. 
Defining the average 
\begin{equation}
\Pi _{-}V_{1}\equiv \lim_{\tau \rightarrow \infty }\frac{1}{\tau }%
\int_{t-\tau }^{t}dsU_{H_{0}}(t,s)V_{1}(s)U_{H_{0}}(s,t),  \label{PiK0}
\end{equation}%
one can show the following property: if $W_1(t)$ is bounded
for negative infinite times, then $\Pi _{-}(V_{1}-D_{1})=0$. This is
satisfied by $D_{1}=\Pi _{-}V_{1}$, the only solution compatible with Eqs. (%
\ref{DD1}) and (\ref{PiK0}). Hence, the averaging $D_{1}=\Pi _{-}V_{1}$ allows one to 
remove secular terms at negative infinite times. 
This gives the precise correspondence between the resonances of stationary problems and the secular terms of time-independent problems.
We remark that
the definition (\ref{PiK0}) of the average  can be in fact recovered from
the formal calculation of the average $\Pi _{K_{0}}V_{1}$ [cf. Eq.~(\ref{D1})]
with respect to $K_{0}=-i\frac{\partial }{\partial t}+H_{0}$ in an extended space, which includes time as a coordinate \cite{SchererIII,art1}.

For a pulsed perturbation that is
  switched on at $t_{i}$ and off at $%
t_{f}$, Eq.~(\ref{PiK0}) becomes \cite{art1} $%
\Pi _{-}V_{1}=U_{H_{0}}(t,t_{i})V(t_{i})U_{H_{0}}(t_{i},t)$. 
This is a
particular solution of Eq. (\ref{DD1}) corresponding to the choice $D_{1}(t_{0})$ $ \equiv $ $
U_{H_{0}}(t_{0},t_{i})V(t_{i})U_{H_{0}}(t_{i},t_{0})$ in Eq. (\ref{DD1(t)}).
An alternate definition of the average: $\Pi _{+}V_{1}\equiv \lim_{\tau
\rightarrow \infty }\frac{1}{\tau }\int_{t}^{t+\tau
}dsU_{H_{0}}(t,s)V_{1}(s)U_{H_{0}}(s,t)$ gives a different averaging $\Pi
_{+}V_{1}=U_{H_{0}}(t,t_{f})V(t_{f})U_{H_{0}}(t_{f},t)$ and  allows one to 
remove secular terms at positive infinite times. 

Generally one cannot remove simultaneously the secular terms at negative and positive large times. 
This shows a conceptual difference between stationary
resonances and secular terms associated with perturbations localized in time. 
Furthermore, it appears that the averaging such as  Eq.~(\ref{PiK0}) is not appropriate, but that a definition which combines the two definitions 
gives a new secular term that could improve the convergence of the algorithm.
This suggests to work with the general solution (\ref{DD1(t)}) of Eq. (\ref%
{DD1}), written with the perturbation evaluated at a free time $t_{1}$ as
the arbitrary operator:%
\begin{equation}
D_{1}(t)=U_{H_{0}}(t,t_{1})V_{1}(t_{1})U_{H_{0}}(t_{1},t).  \label{D1opt}
\end{equation}%
The free $t_{1}$ can then be chosen to minimize the rest after the first iteration, as we describe below.
One has $n$ such free parameters $t_{k},$ $k=1,n$ for $n$ iterations of the
KAM algorithm. There is only one such free parameter for the time-dependent Poincar%
\'{e}-Von Zeipel and Van Vleck methods that are order by order techniques.
An interesting result is that we recover the Magnus expansion from the
time-dependent Poincar\'{e}-Von Zeipel in the particular case of $D_{k}=0$ and $t_{k}^{\prime }=t_{0}$ for $k=1,n$.

{\it Optimization of the perturbation theory}.
After one iteration, the rest $R_{2}(t,t_{0})$
defined in Eq. (\ref{R2}) is associated with a second order operator 
through $R_{2}(t,t_{0})\equiv e^{-i\ep^{2}G_{2}(t)}$ with $G_{2}(t_{0})=0.$
The closer $R_{2}(t,t_{0})$ is to the identity, the smaller the correction terms are, i.e., the more accurate the
approximation is. 
We evaluate the lowest order contribution to $\ep^2 G_{2}(t)$ as
\begin{equation}
\ep^{2}G_{2}^{(2)}(t)=\ep^{2}\int_{t_{0}}^{t}duU_{H_{1}^{\text{e}%
}}(t_{0},u)V_{2}(u)U_{H_{1}^{\text{e}}}(u,t_{0}).
\label{G2}
\end{equation}%
It is this operator that has to remain small for the algorithm to converge.
The size of an operator $A$ can be characterized by the norm $||A||=\sup_{||\psi
||=1}||A\psi ||$ with $\psi $ in the appropriate Hilbert space. 
For an Hermitian matrix this norm reduces to the largest of the absolute values of its eigenvalues.

In order to improve the accuracy we thus seek to minimize $\lambda_2(t)$, the largest of the absolute
values of the eigenvalues of $\ep^2 G_2^{(2)}(t)$, with respect to the free parameters. 
To optimize the KAM algorithm, we have at our disposal two free parameters $%
t_{k}$ and $t_{k}^{\prime }$ at each iteration.
We expect the
parameters $t_{k}$ to significantly affect the convergence, as they are related to secular terms.

{\it Perturbation theory for short intense pulses.}  We consider a
system described by the Hamiltonian $\widehat{H}$ (autonomous or not) and
perturbed by a time-dependent Hamiltonian $\widehat{V}(s)$ whose
characteristic duration is $\tau $. 
The perturbation is assumed to
satisfy $[\widehat{V}(s),\widehat{V}(s_{0})]=0$, $\forall s,s_{0}$ which
is realized in many situations of interest. 
We define a \emph{sudden parameter} $\ep$ as follows.
A dimensionless time $%
t $ and dimensionless operators $H$ and $V(t)$ are defined through $s\equiv \tau t,$ $%
\widehat{H}\equiv \hbar \omega H,$ and $\widehat{V}(s)\equiv \frac{\hbar }{\tau }
V(t)$, leading to the dimensionless Schr\"{o}dinger equation $i\frac{\partial 
}{\partial t}U(t,t_{0})=\left\{ V(t)+\ep H\right\} U(t,t_{0})$, where 
the sudden parameter is defined as $\ep\equiv \omega \tau $. 
We then apply the
perturbation theory described above with the identification $H_{0}(t)\equiv V(t)$ and $V_{1}\equiv H$. This formulation is suited to treat
intense short pulses.

{\it Illustration on a pulsed-driven two-level system}. 

We consider the
case where $H_{0}(t)=\Omega (t)\sigma _{1}$ and $V_{1}=\sigma _{3}$ with $\Omega(t)$ a pulse that is switched on at $t_{i}$ and off at $t_{f}$, and $\sigma _{k}$ the Pauli matrices. 
Notice that, as discussed above, the role of the perturbation and reference Hamiltonian is interchanged.
The pulse area $A\equiv \int_{t_{i}}^{t_{f}}\Omega (u)\,du$ is a
dimensionless parameter that can be fixed independently of the sudden
parameter $\ep$. 
The error between the  numerical solution of the Schr\"{o}dinger equation at the end of the pulse and the result of
 $n$ iterations is defined as $\Delta
_{n}\equiv ||U_{H_1}(t_f,t_i)-U_{H_1}^{(n)}(t_f,t_i)||$. 
We use the pulse shape $\Omega (t)=2A\sin ^{2}\left( \pi t\right) $ for $0\leq
t\leq 1,$ and $0$ elsewhere.

\begin{figure}[h]
\includegraphics[scale=0.7]{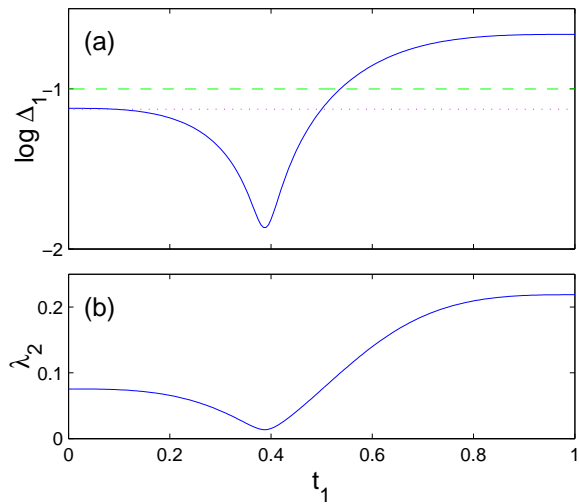}
\caption{(a) Common logarithm of the error $\Delta _{1}$ for the
first order Dyson expansion (dashed line), the first order Magnus expansion (dotted line) and the
first KAM iteration (solid line); and (b)
largest eigenvalue of $\epsilon^2G_{2}^{(2)}(t_{\mathrm{f}})$ as a function of $t_{1}$, for $A=1$, $\ep=0.5$ and $t_{1}^{\prime }=0$.}
\end{figure}

The upper panel of Figure 1 shows a comparison of the  error $\Delta_1$ for the first order Dyson, first order Magnus and
one-iteration KAM methods as a function of $t_{1}$, for a non-perturbative area chosen to produce comparable errors $\Delta_1$ for the Magnus and non-optimized ($t_1=0$) KAM techniques.
The lower panel displays $\lambda _{2}$ the largest of the absolute values of
the eigenvalues of  $\ep^2G_{2}^{(2)}(t_{\mathrm{f}})$
defined in Eq. (\ref{G2}). 
We clearly see that the error of the first KAM iteration is correctly estimated by this eigenvalue $\lambda _{2}$ and, in particular, 
minimized when $\lambda _{2}$ is minimized, i.e., for the value $t_{1}^{\star }$. 
It is worth noting that modifying $t_{1}$ covers more than one order of magnitude in the error, a situation that is not restricted to these values of the parameters.
The optimized solution provides an improvement of the accuracy by almost one order of
magnitude with respect to the Magnus calculation.  

Figure 2 displays a
comparison of the error $\Delta_2$ for the second order Dyson, second order Magnus and two-iteration
KAM methods as a function of $t_{2}$, for $t_{1}=t_{1}^{\star }$. 
It is seen that the Dyson
approach is not applicable in this context of strong field as the second order performs worse than the first one.
\begin{figure}[h]
\includegraphics[scale=0.7]{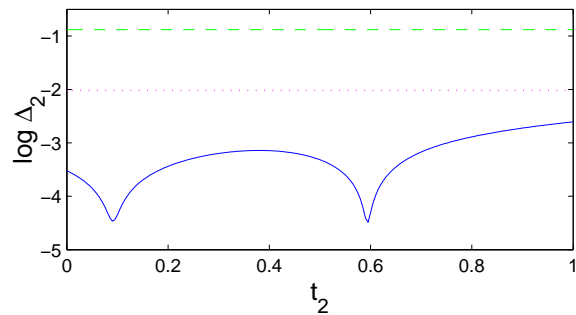}
\caption{Common logarithm of the error $\Delta _{2}$ for the
second order Dyson expansion (dashed line), the second order Magnus expansion (dotted line)
and the second KAM iteration (solid line) with the same parameters as Fig. 1
and $t_{1}=t_{1}^{\star }\approx0.39$.}
\end{figure}
Figure 2 also shows that the second KAM iteration can be 
enhanced by about two orders of magnitude with an appropriate choice of $t_1$ and $t_2$.
This optimized second KAM iteration provides an improvement by two and a half orders of magnitude with respect to the second order
Magnus technique.
Higher iterations of the KAM technique can also be optimized and produce still  better improvement  owing to its superconvergent character.

In conclusion, we have presented an optimized perturbation theory for pulse-driven systems, which applies to a wide class of processes controlled by
intense femtosecond laser pulses. 
The optimization reduces to the evaluation of
eigenvalues and  is therefore easy to implement. 
We anticipate that this approach will be usefull in the context of the laser control of atomic and molecular
processes,
such as the phase space localisation of Rydberg electron \cite{Arbo}, or the alignment
and orientation of molecules \cite{align}.

This research was supported in part by FNRS, ACI \textit{Photonique} and 
 \textit{Conseil R\'{e}gional de Bourgogne}.

\end{document}